\documentclass[letterpaper, english, aps, prl, reprint, twocolumn, superscriptaddress]{revtex4-2}
\usepackage[T1]{fontenc}
\usepackage{color}
\usepackage{amsmath}
\usepackage{amssymb}
\usepackage{graphicx}
\usepackage{multirow}
\usepackage{booktabs}
\usepackage{capt-of}
\usepackage[latin9]{inputenc}
\usepackage{babel}
\usepackage{calc}
\usepackage{float}
\usepackage{dcolumn}

\makeatletter

\usepackage[colorlinks=true, letterpaper=true, pdfstartview=FitV, linkcolor=blue, citecolor=blue, urlcolor=blue]{hyperref}

\renewcommand{\fnum@figure}{Fig. \thefigure}
\renewcommand{\fnum@table}{Table \thetable}

\makeatother

\begin{document}

\title{Universal Magnetic Scaling of the Quantum Metric: Implications for the Magneto-Optical Voigt Effect}

\author{Yongpan Li}
\affiliation{Centre for Quantum Physics, Key Laboratory of Advanced Optoelectronic Quantum Architecture and Measurement (MOE), School of Physics, Beijing Institute of Technology, Beijing, 100081, China}
 
\author{Cheng-Cheng Liu}
\email{ccliu@bit.edu.cn}
\affiliation{Centre for Quantum Physics, Key Laboratory of Advanced Optoelectronic Quantum Architecture and Measurement (MOE), School of Physics, Beijing Institute of Technology, Beijing, 100081, China}

\begin{abstract}
Quantum geometry has emerged as a microscopic framework for understanding a broad range of physical phenomena. The Berry curvature shares the time-reversal-odd axial-vector character of magnetization, naturally allowing linear magnetic scaling in phenomena such as the anomalous Hall and magneto-optical Kerr and Faraday effects. In contrast, the quantum metric is time-reversal even and transforms differently from magnetic order under spatial symmetries, leaving its magnetic scaling less understood. Here we establish universal selection rules governing the magnetic scaling of the quantum metric by combining time reversal and crystalline rotational symmetries. We show that the leading symmetry-allowed magnetic contribution to the in-plane quantum metric is quadratic in the in-plane magnetic order for fourfold- and sixfold-symmetric lattices and for threefold-symmetric lattices with purely in-plane magnetic order. Remarkably, we predict an unconventional linear scaling with a small in-plane magnetic component in threefold-symmetric lattices when the out-of-plane component is large and nearly fixed. Through the symmetric optical conductivity tensor, these scaling laws manifest in the magneto-optical Voigt effect. Furthermore, in the two scaling regimes, both the quantum metric and the Voigt response exhibit distinct rotational behaviors: corotation with the in-plane magnetic order in the quadratic regime and counterrotation by half the magnetic-order rotation angle in the linear regime. Tight-binding calculations verify both scaling regimes, while first-principles calculations for antiferromagnetic bilayer CrI$_3$ exemplify the predicted linear regime. Our selection rules provide a general framework for understanding how magnetic order shapes quantum geometry and constrain the magnetic scaling of a broader class of responses governed by the quantum metric.
\end{abstract}

\maketitle

\textit{Introduction}---Quantum geometry connects the momentum-space structure of Bloch wave functions to a broad range of physical phenomena, including electronic transport, optical absorption, and superconductivity~\cite{ahn_riemannian_2022, gao_field_2014, torma2022superconductivity, peotta2015superfluidity, luhaizhou2024quantume, ezawa2024analytic,yu_quantum_2025, kim_direct_2025, l1hy-6njg, wu2026direct, courtade2026enhanced, zhong2026floquet, li2026anisotropic,bdhy-hnd2, 12cl-b9jj}. The imaginary and real parts of the quantum geometric tensor encode the Berry curvature and the quantum metric, respectively. The Berry curvature shares the time-reversal-odd axial-vector character of magnetization and can therefore exhibit a leading-order linear dependence on magnetization. This linear magnetic scaling can manifest in the anomalous Hall and magneto-optical Kerr and Faraday effects through the intrinsic Hall conductivity and the antisymmetric part of the optical conductivity tensor, respectively~\cite{chang_berry_2008, nagaosa_anomalous_2010, chang_colloquium_2023, kim_determination_2007, tse_giant_2010, kuch_magnetic_2014, feng_large_2015, sivadas_gatecontrollable_2016, higo_large_2018, Han2022Faraday, Ghosh2024QuantumGeometry,f78t-ky69, tornow2026nonlinear}.

In contrast, the quantum metric is time-reversal even and transforms differently from magnetic order under spatial symmetries, making its magnetic scaling far less apparent. Given the important role of the quantum metric in a broad class of physical responses, understanding its magnetic scaling is essential for understanding how these responses depend on magnetic order. In particular, the quantum metric governs the symmetric part of the optical conductivity tensor. Our recent works have identified a quantum metric mechanism for magneto-optical effects in antiferromagnets and altermagnets~\cite{li2025quantum, li2025unconventional}, establishing a microscopic connection between the quantum metric and the magneto-optical Voigt effect. However, the symmetry principles governing the magnetic scaling of the quantum metric remain unclear, leaving the magnetic-order dependence of the Voigt effect and other quantum-metric-related responses unresolved.

\begin{figure}[t]
\centering
\includegraphics[width=1\linewidth]{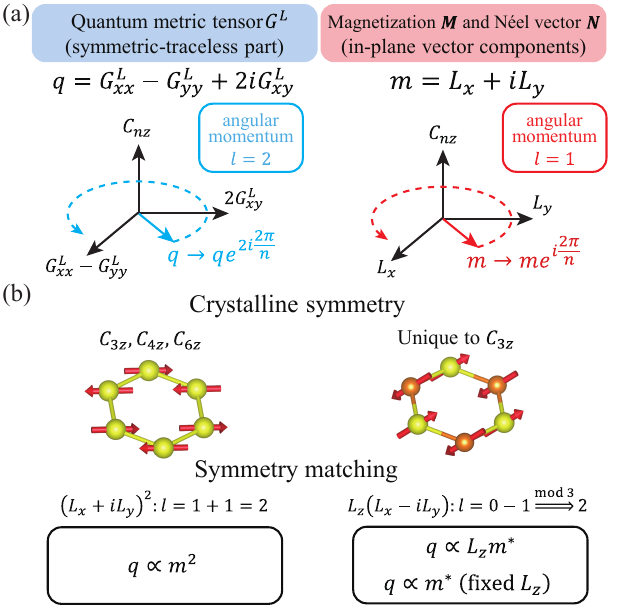}
\caption{Symmetry-based selection rules for the magnetic scaling of the quantum metric. (a) Under crystalline rotations $C_{nz}$, the symmetric-traceless part of the quantum metric tensor, $q=G_{xx}^{\boldsymbol{L}}-G_{yy}^{\boldsymbol{L}}+2iG_{xy}^{\boldsymbol{L}}$, and the in-plane magnetic-order component, $m=L_x+iL_y$ ($\boldsymbol{L}=\boldsymbol{M},\boldsymbol{N}$), carry rotational angular-momentum labels $l=2$ and $l=1$, respectively. (b) Symmetry-allowed magnetic couplings obtained from rotational angular-momentum matching and time-reversal constraints. The quadratic coupling $q\propto m^2$ is allowed by $C_{3z}$, $C_{4z}$, and $C_{6z}$ symmetries, while the mixed coupling $q\propto L_zm^*$ is allowed only by $C_{3z}$ among these symmetries, yielding linear scaling $q\propto m^*$ for a large and nearly fixed $L_z$.
}
\label{fig:quantummetric}
\end{figure}

In this Letter, we establish universal symmetry-based selection rules for the magnetic scaling of the quantum metric by combining rotational angular-momentum matching with time-reversal constraints, as illustrated in Figs.~\ref{fig:quantummetric}(a) and \ref{fig:quantummetric}(b). We show that, in fourfold- and sixfold-symmetric lattices, the leading symmetry-allowed dependence of the in-plane quantum metric on the in-plane magnetic order is quadratic, irrespective of the out-of-plane component. In threefold-symmetric lattices, the scaling is also quadratic for purely in-plane magnetic order, but we predict a linear scaling with a small in-plane component when the out-of-plane component is large and nearly fixed. We illustrate the physical consequences of these selection rules through the magneto-optical Voigt effect. Based on the general magneto-optical theory~\cite{li2025quantum, li2025unconventional}, we derive simplified expressions for the Voigt angles and show that the Voigt effect inherits the magnetic scaling of the quantum metric through the symmetric part of the optical conductivity tensor. Beyond their distinct magnetic scaling laws, the quadratic and linear scalings lead to contrasting rotational behaviors shared by the quantum metric and the Voigt response: corotation with the in-plane magnetic order in the former case and half-angle counterrotation in the latter.
Tight-binding calculations verify both scaling regimes, while first-principles calculations for antiferromagnetic bilayer CrI$_3$ demonstrate the predicted linear regime. Our work deepens our understanding of how magnetic order shapes quantum geometry and, more broadly, provides guidance for predicting and interpreting magnetic-order-dependent responses governed by the quantum metric.

\textit{Magnetic scaling of the quantum metric}---We establish a symmetry-based relation between the magnetic order parameter $\boldsymbol{L}$, representing either the magnetization $\boldsymbol{M}$ or N\'eel vector $\boldsymbol{N}$, and the frequency-resolved quantum metric. 
We consider the Brillouin-zone-integrated quantum metric spectrum, defined as [see Sec. II of the Supplementary Materials (SM)~\cite{SM}]
\begin{equation}\label{eq:integrated_metric}
G_{ij}(\omega,\boldsymbol{L})=\sum_{\substack{n\in\mathrm{occ}\\m\in\mathrm{unocc}}}
\int_{\mathrm{BZ}}d\boldsymbol{k}\ g^{ij}_{mn\boldsymbol{k}}
\delta(\omega-\omega_{mn\boldsymbol{k}}).
\end{equation}
Here, $g_{mn\boldsymbol{k}}^{ij}$ denotes the component of the quantum metric tensor associated with the transition between the $n$th and $m$th states, and $\hbar\omega_{mn\boldsymbol{k}}=\hbar\omega_m-\hbar\omega_n$ is the corresponding transition energy. The dependence of $g_{mn\boldsymbol{k}}^{ij}$ and $\omega_{mn\mathbf{k}}$ on $\boldsymbol{L}$ is left implicit.

Consider a nonmagnetic parent phase with $C_{nz}$ ($n=3,4,6$) symmetry, which enforces an isotropic in-plane quantum metric spectrum, $G_{ij}(\omega,0)=\delta_{ij}G^0(\omega)$ ($i,j=x,y$). When the magnetic order $\boldsymbol{L}$ breaks $C_{nz}$, $G_{xx}(\omega,\boldsymbol{L})$ and $G_{yy}(\omega,\boldsymbol{L})$ need not be equal, and $G_{xy}(\omega,\boldsymbol{L})$ need not vanish. We define the magnetic-order-induced response as $G_{ij}^{\boldsymbol{L}}(\omega)=G_{ij}(\omega,\boldsymbol{L})-\delta_{ij}G^0(\omega)$ and focus on its symmetric traceless part, represented by $q=G_{xx}^{\boldsymbol{L}}-G_{yy}^{\boldsymbol{L}}+2iG_{xy}^{\boldsymbol{L}}$ [See SM~\cite{SM} Sec. I]. Writing the in-plane magnetic order as $m=L_x+iL_y$, the in-plane rotation $C_{nz}$ transforms $q$ and $m$ as $q\rightarrow e^{2i*2\pi/n}q$ and $m\rightarrow e^{i*2\pi/n}m$ [Fig.~\ref{fig:quantummetric}(a)]. Thus, $q$ and $m$ carry rotational angular momenta $l=2$ and $l=1$, respectively, which determine the symmetry-allowed couplings between the quantum metric spectrum and magnetic order.

We consider a power-series expansion of \(q\) in the magnetic-order components about the nonmagnetic state:
\begin{equation}\label{eq:metric_expansion}
q=\sum_{p,r,s\in\mathbb{Z}_{\geq0}}c_{prs}(\omega)m^p(m^*)^rL_z^s.
\end{equation}
Here, $c_{prs}(\omega)$ are material- and frequency-dependent coefficients, and $\mathbb{Z}_{\geq0}$ denotes the set of nonnegative integers. Since $m$, $m^*$, and $L_z$ carry rotational angular-momentum labels $l=1$, $l=-1$, and $l=0$, respectively, the monomial $m^p(m^*)^rL_z^s$ carries $l=p-r$. For this monomial to contribute to $q$, its rotational angular momentum must match that of $q$, i.e., $l=2$ modulo $n$ under $C_{nz}$ symmetry.

Combining this angular-momentum matching condition with the time-reversal-even character of $q$ yields the selection rules
\begin{equation}\label{eq:selet_rule}
p-r=2\pmod n,\quad p+r+s=0\pmod2.
\end{equation}
These selection rules constrain the magnetic-order expansion to all orders, irrespective of the microscopic details of the system. The expansion coefficients $c_{prs}$, however, depend on the electronic structure and frequency and may vanish in the presence of additional symmetries. Unlike the magnetization $\boldsymbol{M}$, the N\'eel vector $\boldsymbol{N}$ may acquire an additional minus sign under $C_{nz}$ if the rotation exchanges the two magnetic sublattices. Nevertheless, because only even-order terms in the magnetic order are allowed by time reversal, this additional sign cancels in every allowed term. The same selection rules therefore apply to both $\boldsymbol{M}$ and $\boldsymbol{N}$.

For $C_{3z}$-, $C_{4z}$-, and $C_{6z}$-symmetric lattices, the quadratic term $m^2$ ($l=2$) is symmetry allowed, while $(m^*)^2$ ($l=-2$) is additionally allowed for $C_{4z}$ because $-2\equiv 2$ (mod 4). Remarkably, $C_{3z}$ symmetry permits an additional coupling between the in-plane and out-of-plane components of magnetic order. The mixed term $L_zm^*$ carries rotational angular momentum $l=-1\equiv 2$ (mod 3) and is even under $\boldsymbol{L}\rightarrow-\boldsymbol{L}$. For $|L_z|\gg|m|$, this mixed coupling is expected to dominate over the quadratic coupling. Consequently, for a large and nearly fixed out-of-plane magnetic component, we predict a leading linear dependence on the in-plane magnetic order,
\begin{equation}
q\propto m^*,\quad (\text{large and fixed}\ L_z).
\end{equation}
For a large and nearly fixed $L_z$, higher-order symmetry-allowed terms of the form $L_z^{2j+1}m^*$ with $j$ a positive integer may also contribute. These terms renormalize the coefficient of the linear coupling without altering its scaling with the in-plane magnetic order. Therefore, in the regime $|L_z|\gg|m|$, the leading symmetry-allowed dependence remains linear in the in-plane magnetic order. Importantly, since neither $l=1$ nor $l=-1$ matches $l=2$ modulo 4 or 6, $C_{4z}$ and $C_{6z}$ symmetries strictly forbid terms linear in the in-plane magnetic order, irrespective of the magnitude of $L_z$. In contrast, $C_{3z}$ symmetry permits such terms through coupling to the out-of-plane magnetic order. These scaling laws are summarized in Fig. ~\ref{fig:quantummetric}(b).

Beyond their distinct magnetic scaling laws, the two couplings give rise to different rotational behaviors of the quantum metric. Writing $q=|q|e^{2i\alpha}$ and $m=|m|e^{i\varphi}$, where $\varphi$ is the azimuth of the in-plane magnetic order, we obtain $\alpha(\varphi)-\alpha(0)=\varphi$ for the quadratic coupling $q\propto m^2$ and $\alpha(\varphi)-\alpha(0)=-\varphi/2$ for the linear coupling $q\propto L_zm^*$. Thus, the quantum metric corotates with the in-plane magnetic order in the former case but counterrotates by half the magnetic-order rotation angle in the latter.

\begin{figure}[t]
\centering
\includegraphics[width=1\linewidth]{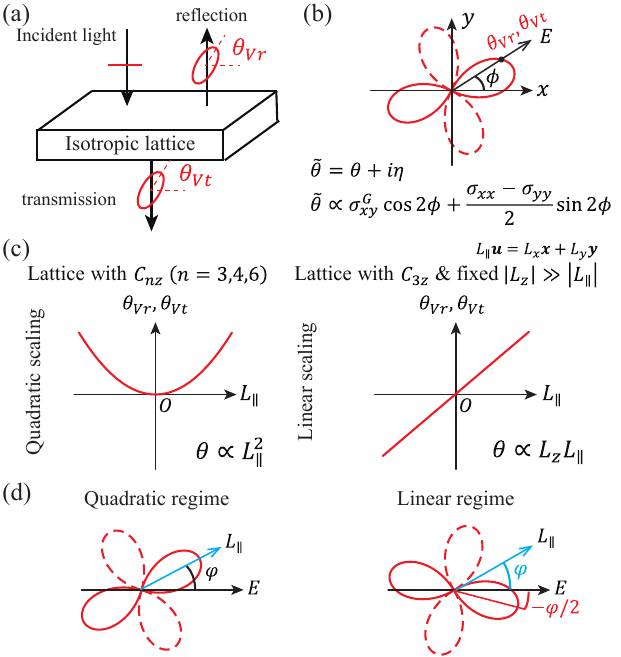}
\caption{Optical manifestations of the magnetic scaling of the quantum metric in the magneto-optical Voigt effect. (a) Measurement geometry for the Voigt effect in reflection and transmission. (b) Dependence of the Voigt rotation on the incident-light polarization direction. The solid and dashed red lobes represent positive and negative Voigt rotations, respectively. (c) Leading symmetry-allowed magnetic scaling of the Voigt rotations $\theta_{Vr}$ and $\theta_{Vt}$. Quadratic scaling is allowed in $C_{4z}$- and $C_{6z}$-symmetric lattices, as well as in $C_{3z}$-symmetric lattices with purely in-plane magnetic order. In $C_{3z}$-symmetric lattices with a large and nearly fixed $L_z$, the mixed coupling gives rise to linear scaling with a small $L_{\parallel}$. (d) Distinct rotational behaviors of the Voigt rotation. Upon rotating the in-plane magnetic order by an angle $\varphi$, the angular pattern of the Voigt rotation corotates by $\varphi$ for the quadratic scaling but counterrotates by $\varphi/2$ for the linear scaling. Blue arrows denote the in-plane magnetic order. The Voigt ellipticity obeys the same magnetic scaling and rotational laws.}
\label{fig:Voigteffect}
\end{figure}

\textit{Application to the magneto-optical Voigt effect}---The magneto-optical Voigt effect describes the rotation and ellipticity of linearly polarized light upon reflection from or transmission through a magnetically ordered medium~\cite{Freiser1968Magnetooptic, Smolenskii1975Birefringence, Schafer1990ReflectionVoigt, Liang2015QuantitativeQMOKE, Saidl2017CuMnAs, Janda2018VoigtMicroscope, Zhao2021Mn3Sn, Wagner2022RuCl3, Schmoranzerova2022YIG, Sepehrinia2024DiracMLD, Wohlrath2025QMOKESpectroscopy, Li2025CrCl3, Marui2026PolarQuadraticKerr, Yang2026MnTe}, as illustrated in Fig. \ref{fig:Voigteffect}(a). Here we consider linearly polarized light incident normally on the sample surface along the $z$ direction. In contrast to the Kerr and Faraday effects, which originate from the antisymmetric part of the optical conductivity tensor, the Voigt effect is governed by its symmetric part~\cite{li2025quantum, li2025unconventional}.

The quantum metric spectrum determines the in-plane symmetric-traceless part of the interband optical conductivity tensor through the following linear spectral relations:
\begin{align}
\sigma_{xy}^{G}(\omega)=&\int_0^\infty d\Omega\,K(\omega,\Omega)G_{xy}^{\boldsymbol{L}}(\Omega),\label{eq:sp1}\\
\sigma_{xx}(\omega)-\sigma_{yy}(\omega)=&\int_0^\infty d\Omega\,K(\omega,\Omega)[G_{xx}^{\boldsymbol{L}}(\Omega)-G_{yy}^{\boldsymbol{L}}(\Omega)]\label{eq:sp2}.
\end{align}
Here, $K$ is a scalar optical-response kernel common to all tensor components [see Sec. II of the SM~\cite{SM}]. Consequently, $(\sigma_{xx}-\sigma_{yy},2\sigma_{xy}^{G})$ and $(G_{xx}^{\boldsymbol L}-G_{yy}^{\boldsymbol L},\,2G_{xy}^{\boldsymbol L})$ share the same transformation properties under spatial symmetries and obey the same magnetic-order selection rules (\ref{eq:selet_rule}).

For the $C_{nz}$-symmetric lattices ($n=3,4,6$) considered here, the nonmagnetic state is optically isotropic in the $xy$ plane. The weak anisotropy introduced by magnetic order implies $|\sigma_{xx}|,|\sigma_{yy}|\gg|\sigma_{xy}^{G}|,|\sigma_{xx}-\sigma_{yy}|$. For three-dimensional (3D) magnets, the general magneto-optical theory~\cite{li2025quantum, li2025unconventional} yields the following simplified expressions for the complex Voigt angles in reflection $\widetilde{\theta}_{Vr}=\theta_{Vr}+i\eta_{Vr}$ and transmission $\widetilde{\theta}_{Vt}=\theta_{Vt}+i\eta_{Vt}$ [see Sec.~III of the SM~\cite{SM}]:
\begin{align}
\widetilde{\theta}_{Vr}^{\mathrm{3D}}=\frac{\sigma_{xy}^{G}}{\sigma_{d}\sqrt{1+i\sigma_{d}/(\epsilon_0\omega)}},\label{eq:voigt_3d}\\
\widetilde{\theta}_{Vt}^{\mathrm{3D}}=-\frac{d\,\sigma_{xy}^{G}}{2\epsilon_0 c \sqrt{1+i\sigma_{d}/(\epsilon_0\omega)}}.
\end{align}
Here, $c$ is the speed of light in vacuum, $d$ is the propagation distance in the medium, and $\sigma_d=(\sigma_{xx}+\sigma_{yy})/2$ denotes the isotropic component of the in-plane optical conductivity and is invariant under rotations of the in-plane coordinate axes. The real and imaginary parts of the complex Voigt angle describe the polarization rotation and ellipticity, respectively.

For two-dimensional (2D) magnets, the Voigt angles also depend on the measurement geometry and the optical properties of the substrate. Considering an optically isotropic, non-absorbing, and nonmagnetic wedge-shaped substrate~\cite{kim_determination_2007}, we simplify the general expressions~\cite{li2025quantum, li2025unconventional} to [see Sec. III of the SM~\cite{SM}]
\begin{align}
\widetilde{\theta}_{Vr}^{\mathrm{2D}}=\frac{2Z_0d\sigma_{xy}^{G}}{(n_s+Z_0d\sigma_{d})^2-1},\\
\widetilde{\theta}_{Vt}^{\mathrm{2D}}=-\frac{Z_0d\sigma_{xy}^{G}}{1+n_s+Z_0d\sigma_{d}},\label{eq:voigt_2d}
\end{align}
where $Z_0=1/(\epsilon_0 c)$ is the vacuum impedance, $n_s$ is the refractive index of the substrate, and $d$ is the thickness of the sample.

Although the simplified expressions for the Voigt angles [Eqs. (\ref{eq:voigt_3d})--(\ref{eq:voigt_2d})] are formally analogous to those for the Kerr and Faraday angles~\cite{kuch_magnetic_2014, kim_determination_2007} , they originate from distinct quantum-geometric contributions. The Voigt effect involves the symmetric transverse optical conductivity arising from the quantum metric, $\sigma_{xy}^{G}=\sigma_{yx}^{G}$, whereas the conventional Kerr and Faraday effects involve the antisymmetric component associated with the Berry curvature, $\sigma_{xy}^{\Omega}=-\sigma_{yx}^{\Omega}$.

Equations~\eqref{eq:voigt_3d}--\eqref{eq:voigt_2d} show that, to leading order, the complex Voigt angles are proportional to $\sigma_{xy}^{G}$. When the incident-light polarization is rotated by $\phi$ [Fig.~\ref{fig:Voigteffect}(b)], the corresponding relation becomes $\tilde{\theta}\propto\frac{1}{2}(\sigma_{xx}-\sigma_{yy})\sin2\phi+\sigma_{xy}^{G}\cos2\phi$. Together with the linear spectral relations [Eqs. \eqref{eq:sp1} and \eqref{eq:sp2}], this establishes that the Voigt response inherits the magnetic scaling of the quantum metric.

We write the magnetic order as $\boldsymbol L=L_\parallel\boldsymbol{u}+L_z\boldsymbol{z}$, where $\boldsymbol{u}$ is an in-plane unit vector and $L_{\parallel}$ and $L_z$ are the signed amplitudes. For $C_{4z}$- and $C_{6z}$-symmetric lattices, as well as $C_{3z}$-symmetric lattices with purely in-plane magnetic order, the leading symmetry-allowed Voigt response is quadratic in $L_{\parallel}$, i.e., $\tilde{\theta}\propto L_{\parallel}^2$. Remarkably, for $C_{3z}$-symmetric lattices with a large and nearly fixed $L_z$ satisfying $|L_z|\gg|L_{\parallel}|$, the mixed coupling $L_zm^*$ gives rise to a leading linear Voigt response,
\begin{equation}\label{eq:Voigt_linear_scaling}
\widetilde{\theta}\propto L_{\parallel}.
\end{equation}
Consequently, at fixed $L_z$ and incident-light polarization, the leading Voigt response reverses sign upon reversal of $L_{\parallel}$, in contrast to the quadratic response. The two magnetic scaling laws are illustrated in Fig.~\ref{fig:Voigteffect}(c).

To derive the explicit dependence of the Voigt angle on the orientation of $L_\parallel\boldsymbol{u}$, let $\varphi$ denote the azimuth of the in-plane magnetic order, with $\boldsymbol{u}=\boldsymbol{x}\cos\varphi+\boldsymbol{y}\sin\varphi$. For a $C_{6z}$-symmetric lattice, or for a $C_{3z}$-symmetric lattice with $L_z=0$, the leading Voigt response arises from the $m^2$ term in $q$, and takes the form [See SM~\cite{SM} Sec. V]
\begin{equation}
\widetilde{\theta}\propto\frac{(\sigma_{xx}-\sigma_{yy})}{2}\sin2(\phi+\varphi)+\sigma_{xy}^{G}\cos 2(\phi+\varphi).\label{eq:voigt_para}
\end{equation}
For a $C_{3z}$-symmetric lattice with a large and fixed out-of-plane component $L_z$ and $|m|\ll|L_z|$, the mixed term $L_zm^*$ gives the leading Voigt response in the form [See SM~\cite{SM} Sec. V]
\begin{equation}
\widetilde{\theta}\propto\frac{\sigma_{xx}-\sigma_{yy}}{2}\sin(2\phi-\varphi)+\sigma_{xy}^{G}\cos(2\phi-\varphi).\label{eq:voigt_linear}
\end{equation}
In a $C_{4z}$-symmetric lattice, both $m^2$ and $(m^*)^2$ are allowed at leading quadratic order. In a $C_{3z}$-symmetric lattice with finite $L_z$, both $m^2$ and $L_zm^*$ can contribute. In either case, the coexistence of two couplings leads to an intricate angular dependence of the Voigt response, for which explicit expressions are derived in Sec.~V of the SM~\cite{SM}.

Equations ~\eqref{eq:voigt_para}--\eqref{eq:voigt_linear} reveal that the distinct rotational behaviors of the quantum metric are inherited by the Voigt response. Specifically, they yield $\tilde{\theta}(\phi,\varphi)=\tilde{\theta}(\phi+\varphi,0)$ and $\tilde{\theta}(\phi,\varphi)=\tilde{\theta}(\phi-\varphi/2,0)$, respectively. As a result, the angular pattern of the Voigt response corotates with the in-plane magnetic order in the quadratic regime but counterrotates by half the magnetic-order rotation angle in the linear regime, as illustrated in Fig. ~\ref{fig:Voigteffect}(d).  This half-angle counterrotation provides a distinct optical signature of the magnetic coupling underlying the linear scaling.

\begin{figure}[b]
\centering
\includegraphics[width=1\linewidth]{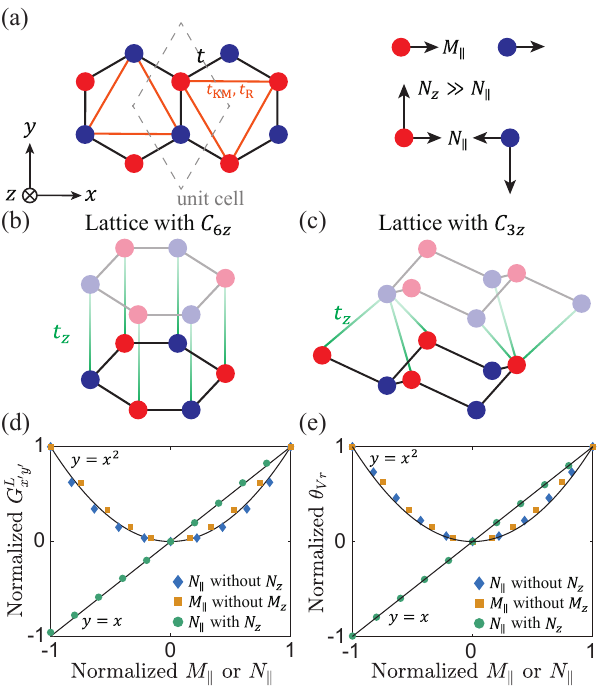}
\caption{Magnetic scaling of the quantum metric and the magneto-optical Voigt effect in tight-binding models. (a) Schematic of the intralayer hopping and magnetic configurations. The magnetic order parameter $\boldsymbol{L}$ ($\boldsymbol{L}=\boldsymbol{M},\boldsymbol{N}$) has an in-plane component $L_{\parallel}$ and an out-of-plane component $L_z$. (b,c) Interlayer hopping geometries for stacked planar and buckled honeycomb layers, respectively. Purely in-plane magnetic order is considered in (b), while a small $N_{\parallel}$ and a large and nearly fixed $N_z$ ($|N_z|\gg|N_{\parallel}|$) are considered in (c). (d,e) Normalized quantum metric $G^{\boldsymbol{L}}_{x'y'}$ and Voigt rotation $\theta_{Vr}$, respectively, as functions of the normalized in-plane magnetic order parameters $M_{\parallel}$ or $N_{\parallel}$. The blue diamonds and orange squares exhibit quadratic scaling for purely in-plane magnetic order, whereas the green circles exhibit linear scaling in the presence of a large and nearly fixed $N_z$. The black curves represent the reference functions $y=x^2$ and $y=x$. The photon energies are $3.5t$ for $M_{\parallel}$ without $M_z$, $6.0t$ for $N_{\parallel}$ without $N_z$, and $14t$ for $N_{\parallel}$ with $N_z$. $t_{\mathrm{KM}}=t_R=0.1t$ and $t_z=0.1t$.
}
\label{fig:tbmodel}
\end{figure}

\textit{Models}---To verify the predicted magnetic scaling laws, we consider tight-binding models on stacked honeycomb lattices with different rotational symmetries, as illustrated in Figs. ~\ref{fig:tbmodel}(a)-(c). The Hamiltonian reads
\begin{align}\label{eq:tb_model}
H=&-t\sum_{\langle ij\rangle}c_i^\dagger c_j+H_z+\sum_i c_i^\dagger [(\boldsymbol{\Delta}_M+\xi_i\boldsymbol{\Delta}_N)\cdot\boldsymbol{s}]c_i\nonumber\\
&+\sum_{\langle\!\langle ij\rangle\!\rangle}c_i^\dagger\left[it_{\mathrm{KM}}\nu_{ij}s_z-it_R\xi_i(\boldsymbol{s}\times\hat{\boldsymbol d}_{ij})_z\right]c_j.
\end{align}
Here, $c_i=(c_{i\uparrow},c_{i\downarrow})^T$, $\boldsymbol{s}=(s_x,s_y,s_z)$ denotes the spin Pauli matrices, and $\xi_i=+1$ ($-1$) on sublattice $A$ ($B$). The first two terms describe intralayer nearest-neighbor and interlayer hoppings, respectively. The exchange fields $\boldsymbol{\Delta}_M$ and $\boldsymbol{\Delta}_N$ couple to uniform magnetization and N\'eel order, respectively. The last term comprises the Kane--Mele spin--orbit coupling (SOC)~\cite{PhysRevLett.95.226801,PhysRevLett.95.146802} and the intrinsic Rashba SOC~\cite{PhysRevLett.107.076802,liu_lowenergy_2011}. $\nu_{ij}=\boldsymbol{d}_i\times\boldsymbol{d}_j/|\boldsymbol{d}_i\times\boldsymbol{d}_j|$, where $\boldsymbol{d}_i$ and $\boldsymbol{d}_j$ are two nearest bonds connecting the next-nearest neighbors $i$ and $j$. $[(\boldsymbol{s}\times\hat{\boldsymbol{d}}_{ij})_z]_{\alpha\beta}=s^x_{\alpha\beta}d_{ij}^y-s^y_{\alpha\beta}d_{ij}^x$, where $\hat{\boldsymbol{d}}_{ij}$ represents the unit vector pointing from site $j$ to $i$.

For the $C_{6z}$-symmetric interlayer hopping geometry shown in Fig.~\ref{fig:tbmodel}(b), the vertically aligned sites on the same sublattice are connected by $H_z=-t_z\sum_{i}c_{i}^\dagger c_{i+z}+\mathrm{H.c.}$. We separately introduce a uniform in-plane exchange field, $\boldsymbol{\Delta}_M=\Delta_\parallel\boldsymbol{x}$ and $\boldsymbol{\Delta}_N=0$, and a staggered in-plane exchange field, $\boldsymbol{\Delta}_M=0$ and $\boldsymbol{\Delta}_N=\Delta_\parallel\boldsymbol{x}$, generating an in-plane magnetization $\boldsymbol{M}_\parallel$ and N\'eel vector $\boldsymbol{N}_\parallel$, respectively. In both cases, the signed exchange-field amplitude $\Delta_\parallel$ is varied from $-0.5t$ to $0.5t$ in steps of $0.1t$. For the $C_{3z}$-symmetric interlayer hopping pattern shown in Fig.~\ref{fig:tbmodel}(c), each $A$ ($B$) site in a given layer is connected to the three nearest $B$ ($A$) sites in the layer above (below) by $H_z=-t_z\sum_{\langle ij\rangle}c_i^\dagger c_{j-\xi_jz}$, with $t_z=t$. We apply a staggered exchange field, $\boldsymbol{\Delta}_M=0$ and $\boldsymbol{\Delta}_N=\Delta_\parallel\boldsymbol{y}+\Delta_z\boldsymbol{z}$. The in-plane component $\Delta_\parallel$ is varied from $-0.05t$ to $0.05t$ in steps of $0.01t$, while the out-of-plane exchange-field component is fixed at $\Delta_z=4t$. This generates a N\'eel vector with a tunable small in-plane component $\boldsymbol{N}_\parallel$ and a large and nearly fixed out-of-plane component $N_z$, satisfying $|N_z|\gg|N_\parallel|$.

Since all magnetic states preserve either the $M_x$ or the $M_y$ mirror symmetry, the Voigt angle reaches the maximum magnitude when the incident-light polarization direction is at $45^\circ$ to the $x$ axis [See SM~\cite{SM} Sec.~IV]. We choose the rotated coordinate axes  $\boldsymbol{x}'=(\boldsymbol{x}-\boldsymbol{y})/\sqrt{2}$ and $\boldsymbol{y}'=(\boldsymbol{x}+\boldsymbol{y})/\sqrt{2}$, with the incident light polarized along $\boldsymbol{x}'$. Figures~\ref{fig:tbmodel}(d) and \ref{fig:tbmodel}(e) show the calculated quantum metric $G^{\boldsymbol{L}}_{x'y'}$ and Voigt rotation $\theta_{Vr}$ as functions of the signed in-plane magnetic order parameters $M_\parallel$ or $N_\parallel$. We define $M_\parallel=(\langle \boldsymbol{s}_\parallel\rangle_A+\langle \boldsymbol{s}_\parallel\rangle_B)/2$ and $N_\parallel=(\langle \boldsymbol{s}_\parallel\rangle_A-\langle\boldsymbol{s}_\parallel\rangle_B)/2$, where $\langle \boldsymbol{s}_\parallel\rangle_{A(B)}$ denotes the expectation value of $\boldsymbol{s}_\parallel=\boldsymbol{s}_x+\boldsymbol{s}_y$ on sublattice A (B). Each data set is normalized to its maximum absolute value to facilitate comparison between different magnetic configurations. 

For purely in-plane magnetic order, both $G^{\boldsymbol{L}}_{x'y'}$ and $\theta_{Vr}$ exhibit quadratic scaling with either $M_{\parallel}$ (orange squares) or $N_{\parallel}$ (blue diamonds), consistent with our symmetry-based selection rules. This behavior also agrees with the commonly observed quadratic dependence of the Voigt effect on magnetization in ferromagnets~\cite{PhysRevB.89.085203,PhysRevLett.87.047401,Tes2012,Hamrlova2013Quadratic, Liang2015QuantitativeQMOKE,PhysRevLett.94.227203,Hamrle_2007} or on the sublattice magnetic-moment magnitude in antiferromagnets~\cite{Saidl2017CuMnAs,Zhao2021Mn3Sn,Zhao2021QuadraticMO, Nemec2018Antiferromagnetic,Wagner2022RuCl3, PhysRevApplied.16.014037,PhysRevB.105.224428}.

In contrast, for the model with a $C_{3z}$-symmetric nonmagnetic reference state, a large and nearly fixed out-of-plane N\'eel-vector component $N_z$ leads to a distinct magnetic scaling. Both $G^{\boldsymbol{L}}_{x'y'}$ and $\theta_{Vr}$ vary linearly with $N_{\parallel}$, as indicated by the green circles in Figs.~\ref{fig:tbmodel}(d) and \ref{fig:tbmodel}(e), and reverse sign upon reversal of $N_{\parallel}$ at fixed $N_z$. These results verify the predicted leading linear scaling of the quantum metric and its manifestation in the Voigt effect.

\begin{figure}[t]
\centering
\includegraphics[width=1\linewidth]{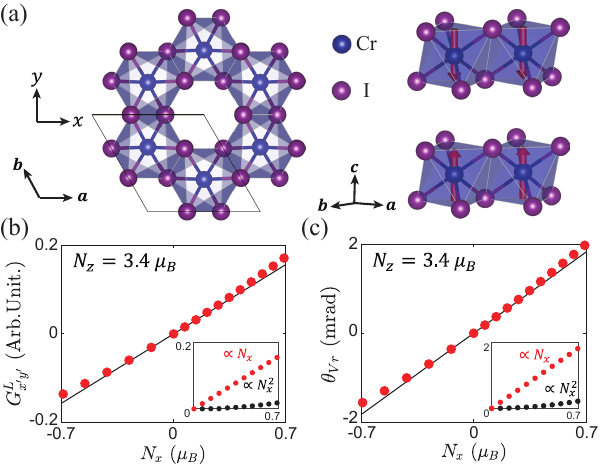}
\caption{Linear magnetic scaling of the quantum metric and the magneto-optical Voigt effect exemplified by antiferromagnetic bilayer CrI$_3$. (a) Schematic of the crystal structure and magnetic configuration of antiferromagnetic bilayer CrI$_3$, whose equilibrium N\'eel vector is oriented along the $c$ axis in the considered configuration. (b,c) Calculated quantum metric $G^{\boldsymbol{L}}_{x'y'}$ and Voigt rotation $\theta_{Vr}$, respectively, as functions of the in-plane N\'eel-vector component $N_x$ at a photon energy of $2.0$ eV, with a large and nearly fixed out-of-plane component $N_z\simeq3.4\mu_B$. Both quantities exhibit approximately linear scaling and sign reversal upon reversal of $N_x$. The insets show the decomposition into linear and higher-order contributions, with the linear contribution being dominant.}
\label{fig:CrI3}
\end{figure}

\textit{Materials}---As a material example of our general predictions, we investigate the linear magnetic scaling of the quantum metric and the magneto-optical Voigt effect in antiferromagnetic bilayer CrI$_3$. We consider the $C_{3z}$-symmetric bilayer structure with intralayer ferromagnetic order and interlayer antiferromagnetic order~\cite{huang2017layer} as illustrated in Fig.~\ref{fig:CrI3}(a). Owing to space-time inversion symmetry, the Berry curvature vanishes and the associated antisymmetric optical conductivity is forbidden. When the N\'eel vector is aligned along the $c$ axis, the magnetic state additionally preserves $C_{3z}$ symmetry, enforcing the Voigt rotation to vanish.

We introduce a small in-plane N\'eel-vector component $N_x$ while maintaining the large out-of-plane component nearly fixed at $N_z\simeq3.4\mu_B$, as obtained from first-principles calculations. For $N_x\neq0$, the system preserves space-time inversion symmetry and mirror symmetry $M_y$ but breaks $C_{3z}$ symmetry. Our selection rules allow both the quadratic coupling $N_x^2$ and the mixed coupling $N_xN_z$ in the magnetic-order dependence of the quantum metric. For $|N_x|\ll|N_z|$ at nearly fixed $N_z$, the mixed coupling is expected to dominate, giving rise to a leading linear dependence of the quantum metric and the Voigt rotation on $N_x$.

To examine the Voigt rotation with the maximum magnitude, we consider the rotated coordinate system $(\boldsymbol{x}',\boldsymbol{y}')=(\boldsymbol{x}-\boldsymbol{y},\boldsymbol{x}+\boldsymbol{y})/\sqrt{2}$, with the incident light polarized along $\boldsymbol{x}'$. 
Figures~\ref{fig:CrI3}(b) and \ref{fig:CrI3}(c) show the calculated quantum metric $G^{\boldsymbol{L}}_{x'y'}$ and Voigt rotation $\theta_{Vr}$, respectively, as functions of $N_x$. At nearly fixed $N_z$, both quantities exhibit approximately linear scaling and reverse sign upon reversal of $N_x$. The linear scaling becomes increasingly accurate as $|N_x|$ decreases, consistent with the leading symmetry-allowed contribution. This sign reversal at fixed $N_z$ is consistent with the time-reversal-even nature of the response, since the mixed coupling $N_xN_z$ remains invariant under simultaneous reversal of both N\'eel-vector components. The decomposition into linear and higher-order contributions [insets of Figs.~\ref{fig:CrI3}(b) and \ref{fig:CrI3}(c)] reveals the dominance of the linear contribution, further corroborating our symmetry-based prediction.

\textit{Discussion}---An important implication of our selection rules is that the time-reversal-even character of the quantum metric does not preclude a linear dependence on an individual component of magnetic order. For the mixed coupling allowed by threefold rotational symmetry, the sign reversal and half-angle counterrotation of the Voigt response upon reversal and reorientation of the in-plane magnetic order, respectively, provide complementary optical signatures of the underlying quantum metric coupling.

Using available experimental techniques, our predicted linear scaling can be experimentally tested in both conventional antiferromagnets and altermagnets. Magnetic-field-induced reorientation~\cite{Lebrun2020,Kriegner2017}, optical excitation of coherent spin waves~\cite{Satoh2010,Zhang2020}, and electrical control via spin--orbit torque~\cite{Wadley2016,Xue2021} offer potential routes to tuning a small in-plane magnetic component $N_{\parallel}$ while maintaining a large and nearly fixed $N_z$. At fixed photon energy and incident-light polarization, measuring the Voigt rotation as a function of $N_{\parallel}$ would directly test the predicted linear scaling $\theta_{Vr}\propto N_{\parallel}$. Furthermore, comparing the responses for opposite $N_{\parallel}$ at fixed $N_z$ would separate contributions odd and even in the in-plane magnetic order, allowing the leading linear contribution to be distinguished from the quadratic one in the small-$N_{\parallel}$ regime. These measurements also provide a route to optically determining the N\'eel-vector orientation: with the sign of $N_{\parallel}$ known and the optical-response coefficient calibrated, the mixed coupling between $N_z$ and $N_{\parallel}$ allows the sign of $N_z$ to be inferred from the measured Voigt rotation.

Our selection rules treat the magnetization and N\'eel vector on the same footing. In ferromagnets, a particularly interesting consequence is the distinction between the predicted linear Voigt effect and the linear Kerr and Faraday effects. For light propagating along the rotation axis, the former is governed by the time-reversal-even quantum metric and scales with the in-plane magnetization at a large and nearly fixed out-of-plane magnetization, whereas the latter originate from the time-reversal-odd Berry curvature and scale with the out-of-plane magnetization. More broadly, our selection rules provide a symmetry-based guiding principle for predicting the magnetic-order dependence of quantum-geometric responses without relying on material-specific microscopic models.

\textit{Acknowledgements}---We thank Xu Yang and Zhao-Hua Cheng for valuable discussions. The work is supported by the Science Fund for Creative Research Groups of NSFC (Grant No. 12321004), the NSF of China (Grant No. 12374055), and the Fundamental Research Funds for the Central Universities (Grant No. 2026CX01001).

\bibliography{reference}

\end{document}